\documentclass{proc}
\usepackage[utf8]{inputenc}
\usepackage[hyphens]{url}
\usepackage{amsmath}
\usepackage{graphicx}
\usepackage{caption}
\usepackage{subcaption}
\usepackage{multicol}
\usepackage{multirow}
\usepackage{flushend}
\usepackage{authblk}
\usepackage{hyperref}

\newcommand{\gbar}{\bar{\gamma}}
\newcommand{\thebar}{\bar{\theta}}
\title{Fast and Accurate Forecasting of COVID-19 Deaths\\Using the SIkJ$\alpha$ Model}
\author{Ajitesh Srivastava}
\author{Tianjian Xu}
\author{Viktor K. Prasanna}
\affil{University of Southern California}
\affil{\{ajiteshs,frostxu,prasanna\}@usc.edu}

\begin{document}

\maketitle

\begin{abstract}
    Forecasting the effect of COVID-19 is essential to  design policies that may prepare us to handle the pandemic. Many methods have already been proposed, particularly, to forecast reported cases and deaths at country-level and state-level. Many of these methods are based on traditional epidemiological model which rely on simulations or Bayesian inference to simultaneously learn many parameters at a time. This makes them prone to over-fitting and slow execution. We propose an extension to our model SIkJ$\alpha$ to forecast deaths and show that it can consider the effect of many complexities of the epidemic process and yet be simplified to a few parameters that are learned using fast linear regressions.  We also present an evaluation of our method against seven approaches currently being used by the CDC, based on their two weeks forecast at various times during the pandemic. We demonstrate that our method achieves better root mean squared error compared to these seven approaches during majority of the evaluation period. Further, on a 2 core desktop machine, our approach takes only 3.18s to tune hyper-parameters, learn parameters and generate 100 days of forecasts of reported cases and deaths for all the states in the US. The total execution time for 184 countries is 11.83s and for all the US counties ($>$ 3000) is 101.03s.
\end{abstract}

\section{Introduction}

COVID-19 has severely affected global health and economy. Accurate forecasts of the effect of such a pandemic is essential to design policies regarding resource management and economic decisions. Center for Disease Control is using around 20 different death forecasting approaches for this purpose, including our method~\cite{CDCforecasts}. Many of these models are based on the traditional SEIR model~\cite{li1999global}, and yet they produce different results. This suggests that the choice of the training approach is critical in generating accurate forecasts. Further, fast training and forecasting is desirable so that the policy-makers can generate forecasts for different future scenarios for many regions (countries, states, counties) to choose policies and evaluate possible effects. 

We extend our prior model SIkJ$\alpha$~\cite{srivastava2020learning} to forecast deaths and demonstrate that it provides accurate and fast predictions. 
Often epidemic models are trained through numerical solutions to differential equations~\cite{cintron2020estimation} or through Bayesian inference~\cite{lekone2006statistical,dukic2012tracking}. These training approaches may be prone to overfitting as they attempt to simultaneously learn many parameters of a non-linear model without providing learnability guarantees, with the exception of few works~\cite{srivastava2020data,ducrot2020identifying,magal2018parameter}. 
Instead, we transform our model into a system of simple linear models for learning the parameters.
We show that our model considers many complexities such as unreported cases due to any reason (asymptomatic, mild symptoms, willingness to get tested), immunity (if any) or complete isolation, and reporting delay, and yet, it can be reduced to a system of two linear equations which can be fitted one after the other resulting in fast yet reliable forecasts.
We compare our approach with 7 other models that are being used for forecasting deaths in the United States by the CDC~\cite{CDCforecasts}. We show that our approach outperforms all the other methods during majority of the evaluation period (May 10, 2020 -- June 28, 2020).  On a 2 core desktop machine, our approach takes only 3.18s to tune hyper-parameters, learn parameters and generate 100 days of forecasts of reported cases and deaths for all the states in the US. The total execution time for 184 countries is 11.83s and for all the US counties ($>$ 3000) is 101.03s.

Our US state-level forecasts and global country-level forecasts can be found on our interactive web-page\footnote{\url{https://scc-usc.github.io/ReCOVER-COVID-19/}}, which is updated 3-4 times a week. All the codes for generating the forecasts and performing comparisons with other methods are available on github\footnote{\url{https://github.com/scc-usc/ReCOVER-COVID-19}}.

\section{The SIkJ$\alpha$ Model: Extension and Simplification}
In~\cite{srivastava2020learning}, we proposed the SI-kJ$\alpha$ model %(Fig.~\ref{fig1}) 
for the spread of a virus like COVID-19 across the world which captures (i) temporally varying infection rates (ii) arbitrary regions, and (iii) human mobility patterns. Within every region (hospital/city/state/country), an individual can exist in either one of two states: susceptible and infected. A susceptible individual gets infected when in contact with an infected individual at a rate depending on when that individual got infected, i.e., rate of infection  is $\beta_1$ for an individual infected between $t-1$ and $t-J$, $\beta_2$ for an individual infected between $t-J$ and $t-2J$, and so on, thus resulting in $k$ sub-states of infection.
$J$ is a hyperparameters introduced for a smoothing effect to deal with noisy data.
It also avoids over-fitting the model by using a small $k$ to capture dependency on the last $kJ$ days.
The hypothesis is that how actively one passes on the infection is affected by when they get infected. 
We assume that after being infected for a certain time, individuals no longer spread the infection, i.e., $\exists k$, such that $\beta_i = 0 \forall i>k$.

Also, people traveling from other regions can increase the number of infections in a given region. Suppose $F(q, p)$ represents mobility from region $q$ to region $p$. Further extensions were proposed in ~\cite{srivastava2020data} to account for the fraction of the population that is either immune or completely isolated, thus not participating in the epidemic (non-carriers). The probability of reporting a case is also accounted for. A case may not have been reported if a person is asymptomatic or has mild symptoms, or has symptoms but decides not to get tested.
%This includes estimates of (a) number of people flying from region $q$ to region $p$, and (b) number of people traveling between the regions by road.
Our model is represented by the following system of recurrence relations for a region $p$ at time $t$.

\begin{align}
S_t^{(p)} &= (1-\rho^{(p)})N^{(p)} - I_t^{(p)}, \label{eqn:susceptible}\\
R_t^{(p)} &= \gamma^{(p)} I_{t-\lambda}^{(p)}, \label{eqn:reported}  \\
\Delta I_t^{(p)} &= \frac{S_{t}^{(p)}}{(1-\rho^{(p)}) N^{(p)}} \sum_{i=1}^k \beta_i^{(p)} (I_{t-(i-1)J}^{(p)} -I_{t-iJ}^{(p)}) \nonumber \\
&+\delta \sum_q F(q, p) \frac{\sum_{i=1}^k \beta_i^{(q)} (I_{t-(i-1)J}^{(q)} -I_{t-iJ}^{(q)})}{(1-\rho^{(q)})N^{(q)}}\,,\label{eqn:new_infections}
\end{align}

Equation~\ref{eqn:susceptible} suggests that the number of susceptible individuals $S^{(p)}_t$ is determined by the difference between the population that is not immune/isolated and cumulative infections $I_t$ so far. Here $\rho^{(p)}$ represents the fraction of population out of total population $N^{(p)}$ that is immune/isolated.
Equation~\ref{eqn:reported} suggests that an infected case is reported with probability $\gamma^{(p)}$ after a constant lag $\lambda$, resulting in $R_t^{(p)}$ observed reported cases.
Equation~\ref{eqn:new_infections} describes how new infections $\Delta I_t^{(p)}$ are created after time $t$. It consists of infections from the same region (community spread) accounting for heterogeneous infection rates $\beta_i^{(p)}$. The new infections may also be created by incoming travel (travel spread).
Parameter $\delta$ captures the influence of passengers coming into the region. Observe that the the new infections at time $t$ are dependent only on the new infections in the last $kJ$ time units. Therefore, the model implicitly assumes that after $kJ$ days an individual is not infectious, for any reason such as recovery, death or being quarantined.

\subsection{Death Modeling and Simplification}

We extend this model to measure number of deaths as a function of infected cases:
\begin{equation}\label{eqn:deaths}
    \Delta D_t = \sum_{i=1}^{k_D} \theta_i^{(p)} (I_{t-(i-1)J_D}^{(p)} -I_{t-iJ_D}^{(p)}).
\end{equation}
Equation~\ref{eqn:deaths} suggests that new deaths $\Delta D_t$ occur at time $t$ as individuals infected between time $t-(i-1)J_D$ and $t-iJ_D$ with probability $\theta_i^{(p)}$. Note that we do not require $k_D = k$ or $J_D = J$. For state-level forecasts when many states have experienced a rapid rise of the pandemic, travel is being avoided. Therefore, we may assume that the community spread is dominant and travel spread can be ignored. Ignoring travel spread in Equation~\ref{eqn:new_infections} and combining with ~\ref{eqn:susceptible}, we get
\begin{align}
    \Delta I_t^{(p)} &= \left( 1 - \frac{I_t^{(p)}}{(1-\rho^{(p)}) N^{(p)}} \right) \sum_{i=1}^k \beta_i^{(p)} (I_{t-(i-1)J}^{(p)} -I_{t-iJ}^{(p)})
\end{align}
Combining the above with Equation~\ref{eqn:reported} and setting $\gbar^{(p)} = \gamma^{(p)} (1-\rho^{(p)})$, we get
\begin{equation}\label{eqn:report_final}
    \Delta R_t^{(p)} = \left( 1 - \frac{R_t^{(p)}}{\gbar^{(p)} N^{(p)}} \right) \sum_{i=1}^k \beta_i^{(p)} (R_{t-(i-1)J}^{(p)} -R_{t-iJ}^{(p)})
\end{equation}
Note that the constant lag $\lambda$ has no effect above. Similarly, from Equation~\ref{eqn:deaths} and Equation~\ref{eqn:reported}, and setting $\thebar_i^{(p)} = \theta_i^{(p)} / \gamma_i^{(p)}$, we get
\begin{equation}\label{eqn:death_semifinal}
    \Delta D_t = \sum_{i=1}^{k_D} \thebar_i^{(p)} (R_{t+\lambda-(i-1)J_D}^{(p)} -R_{t+\lambda-iJ_D}^{(p)}).
\end{equation}
Assuming that the time period between contracting the virus and reporting of the positive case is shorter than contracting the virus and death, we can rewrite Equation~\ref{eqn:death_semifinal} as:
\begin{equation}\label{eqn:death_final}
    \Delta D_t = \sum_{i=1}^{k_D} \thebar_i^{(p)} (R_{t-(i-1)J_D}^{(p)} -R_{t-iJ_D}^{(p)}).
\end{equation}
We first generate forecasts for reported cases using Equation~\ref{eqn:report_final} and then use those forecasts along with historical reported cases as inputs in Equation~\ref{eqn:death_final} to forecast deaths.

\begin{figure}[!ht]
    \centering
    \includegraphics[width=0.9\linewidth]{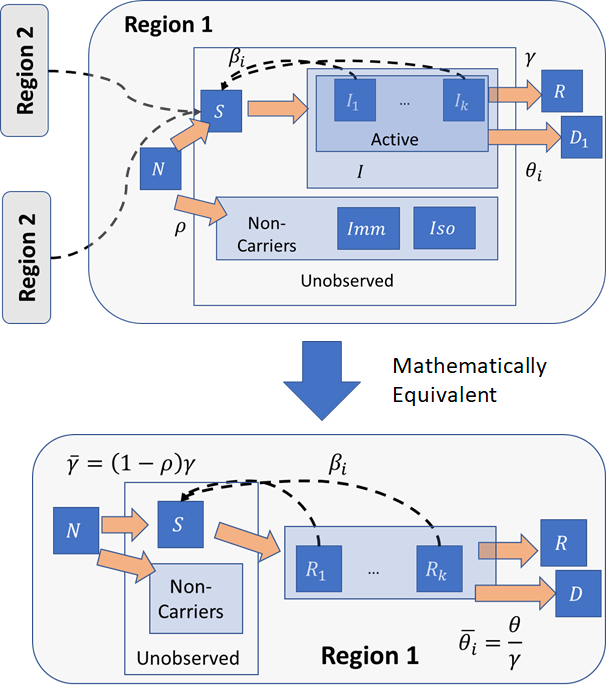}
    \caption{Disregarding travel spread and simplifying the model to reduce the number of parameters. }
    \label{fig:model}
\end{figure}
Figure~\ref{fig:model} summarizes the transformation of the model that we started with and its simplification.

The learning of parameters is described next.

\subsection{Parameter Learning}

We consider $\beta_i^{(p)}$ and $\thebar_i^{(p)}$ as learnable parameters, and $k, k_D, J, J_D$ and $\gbar^{(p)}$ are treated as hyper-parameters. We let each region have independent hyper-parameters. The reason for considering $\gbar^{(p)}$ as a hyper-parameter is to keep the model linear during training. We have tested the model in a non-linear setting and it produces worse results. Further, these parameters may not provide meaningful insights in a non-linear setting -- as shown in~\cite{srivastava2020data}, learning the true value of $\gbar^{(p)}$ can be done reliably only under certain condition. We used the results obtained in~\cite{srivastava2020data} for some US states ($\gbar \in (1/40, 1/20)$) and assume that all states will stay in the same range. 
We observed that $\gbar = 1/40$ generally produced good results, however, the difference in short-term forecasts was not significant when changing $\gbar$. We perform a grid search for $k \in \{1, 2\}$ and $J \in \{7, 8, \dots, 14\}$, while ensuring that $kJ \leq 14$. This is along the lines of the motivation for 14 days of quarantine\footnote{\url{https://www.cdc.gov/coronavirus/2019-ncov/travelers/after-travel-precautions.html}}. The value of $J$ below 7 was not considered driven by the  observed weekly periodicity in the data~\cite{ricon2020seven}, suggesting that we should smooth over at least 7 days. These hyper-parameters are allowed to be different for different states. The values of $k_D$ and $J_D$ can also be selected by a grid search. However $k_D J_D$ is allowed to be higher. 
We observed that $k_D = 2$ and $J_D = 7$ generally performed well. Note that the evaluations to identify best hyper-parameters were performed on a held-out portion of training data (validation set), and no part of test data is used. For each state, the evaluation on the validation set was performed using the root mean squared error (RMSE) of the predictions on the last $H = 7$ days of training data.
\begin{equation}
    RMSE =  \sqrt{\frac{\sum_{t=T-H+1}^T (\Delta \hat{R}_t^{(p)} - \Delta R_t^{(p)})^2}{H}}\,,
\end{equation}
where $\hat{R}_t^{(p)}$ is the observed value of reported cases for state $p$ at time $t$.

We learn the parameters $\beta_i^{(p)}$ as described in~\cite{srivastava2020learning}, which incorporates the fast evolving trend of COVID-19 due to changing policies using weighted least squares. The best fit in the least-squares sense minimizes the sum of squared weighted residuals, i.e., the difference between observed data and predicted values provided by our learned model. The weighing scheme is determined by a hyper-parameter $\alpha \leq 1$ to put more weight to the recent infection trend when learning the model. Lower $\alpha$ implies more emphasis on the more recent data. We minimize the sum of squares of weighted errors as
\begin{align}
    LSE_R = \sum_{t=1}^{T} (\alpha^{T-t}(\Delta \hat{R}_t^{(p)} - \Delta R_t^{(p)}))^2\,
\end{align}
Where $R_t$ is given by the linear equation~\ref{eqn:report_final}. For learning $\thebar_i^{(p)}$, we assume that the death rates do not evolve very rapidly and can be learned considering a window of the last $w$ time steps. This introduces another hyper-parameter, which we find to generally perform well at $w=50$.  We minimize the unweighted least squared error to obtain $\thebar_i^{(p)}$. 
\begin{equation}
    LSE_D = \sum_{t=T-w+1}^{T} (\Delta \hat{D}_t^{(p)} - \Delta D_t^{(p)})^2\,
\end{equation}
where $\hat{D}_t^{(p)}$ is the true number of new deaths at time $t$ in region $p$. Note that before we attempt to learn the parameters, we perform a moving average smoothing over 7 days to reduce noise due to periodicity in reporting. This is performed in irrespective of the smoothing hyper-parameter $J$.

\section{Other Models as Baselines}
We picked those methods as baselines from COVID-19 Forecast Hub\footnote{\url{https://github.com/reichlab/covid19-forecast-hub/tree/master/data-processed}} that had their reports available every Sunday in the month of May and June. This day was chosen based on CDC's definition of a week~\cite{CDCweek}. Here we provide an overview of the baselines.

\paragraph{CU\_SELECT.}
The Shaman Group from Columbia university uses a meta-population county-level SEIR model~\cite{pei2020initial} to make COVID-19 projections in the US based on various contact rates between counties. SELECT is a selection of their forecast reports that tries to fit the realistic contact rates during different time periods given current observations and planned intervention policies. The method considers county-to-county commuters and random visitors, and formulates the transmission during day-time and night-time separately. They use a combination of iterated filtering (IF) and the Ensemble Adjustment Kalman Filter (EAKF)~\cite{anderson2001ensemble} framework \cite{ionides2006inference,king2008inapparent} to calibrate the model's parameters. They integrate their model equations using a Poisson process to represent the stochasticity of the transmission process.

\paragraph{JHU\_IDD.}

The Johns Hopkins ID Dynamics COVID-19 Group has developed a scenario modeling pipeline~\cite{Lemaitre2020.06.11.20127894} for the epidemic projection. The pipeline contains three module: 1) module 1 is an epidemic seeding module that incorporates the first case appearance in data and air travel import model; 2) module 2 is a disease transmission model that takes in the seeding information and simulates a county-level meta-population model with stochastic SEIR disease dynamics; and 3) module 3 is a health outcomes projection model that takes in consideration of realistic time delays between infection, symptoms, hospitalization, intensive care, ventilator use, and death to model deaths and hospitalizations in the population.  As pointed out by the authors~\cite{Lemaitre2020.06.11.20127894}, the model does not account for asymptomatic transmissions.

\paragraph{UCLA\_SuEIR.}

UCLA Statistical Machine Learning Lab\footnote{\url{https://covid19.uclaml.org/}} has proposed a variant of the SEIR model called SuSEIR ~\cite{zou2020epidemic}. It accounts for the untested and unreported COVID-19 cases. The model can be described by the following ODE:
\begin{equation}
        \begin{aligned}
            \frac{dS_{t}}{dt} &= - \frac{\beta (I_{t} + E_{t})S_{t}}{N}, \\
            \frac{dE_{t}}{dt} &= \frac{\beta(I_{t} + E_{t})S_{t}}{N} - \sigma E_{t}, \\
            \frac{dI_{t}}{dt} &= \mu \sigma E_{t} - \gamma I_{t}, \\
            \frac{dR_{t}}{dt} &= \gamma I_{t}
        \end{aligned}
\end{equation}

The key claim of the SuEIR model is that it can infer the untested cases as well as unreported cases, although a discussion on learnability~\cite{srivastava2020data} is not provided. More specifically, they treat the ``Exposed” group in the SEIR model as the individuals that have already been infected and have not been tested, which are also capable of infecting the ``Susceptible” group. Moreover, some of the people in the ``Exposed” group can get tested and transferred to the ``Infectious” group (which are reported to the public), while the rest of them will recover/die and transit to the so-called ``Unreported Recovered” group (which are not reported to the public). They use gradient-based optimizers to minimizing a loss function composed of squared errors on infected and removed cases.

\paragraph{IowaStateLW\_STEM.}

Lily Wang's Research Group from Iowa State University\footnote{\url{https://covid19.stat.iastate.edu}} has developed a non-parametric spatio-temporal epidemic transmission model~\cite{wang2020spatiotemporal}. They have designed a space-time epidemic modeling framework by including area-level characteristics to the traditional SIR model. The model assumes that there are local characteristics of a given area that will not vary with time. It further assumes that such characteristics will influence the daily new cases in the local area and its surrounding areas.

\paragraph{Covid19Sim\_Simulator.}

COVID-19 Simulator is developed to simulate the trajectory of COVID-19 in US state-level by researchers from Mass General Hospital, Harvard Medical School, Georgia Tech and Boston Medical Center\footnote{\url{https://covid19sim.org/}}. The compartment model is defined using the SEIR model with continuous time progression. 
The model parameters are learned by first defining sensible ranges on the parameters and then using simulated annealing\footnote{\url{https://covid19sim.org/images/docs/COVID-19_simulator_methodology_download_20200507.pdf}}.
The model can simulate three intervention strategies: minimal restriction, current intervention in each state, and complete lockdown. The forecast reports we use is the simulation of the current intervention in each state.

\paragraph{CovidActNow.}

CovidActNow is a team of computer scientists, epidemiologists, health experts, and public policy leaders\footnote{\url{https://covidactnow.org/}}. They use an SEIR model to forecast trajectory of COVID-19 in US state-level. Their forecasts rely on fitting predicted cases, deaths, and hospitalizations to the observations using known ranges of the parameters and maximum-likelihood optimization.

\paragraph{YYG\_ParamSearch.}
YYG\_ParamSearch uses an SEIR model\footnote{\url{https://covid19-projections.com/}}. It fixes some of the parameters that it considers to be known such as the latency of reporting, infectious period, time between illness onset to hospitalization, time between illness onset to death, hospital stay time, and time to recovery. It learns the basic reproduction number $R_0$, mortality rate, and the reproduction numbers in various mitigation scenarios using a grid-search, evaluated using means squared error.

\section{Experiments}

\subsection{Data}
\begin{figure*}[!htbp]
\centering
\begin{subfigure}{\textwidth}
\includegraphics[width=\linewidth]{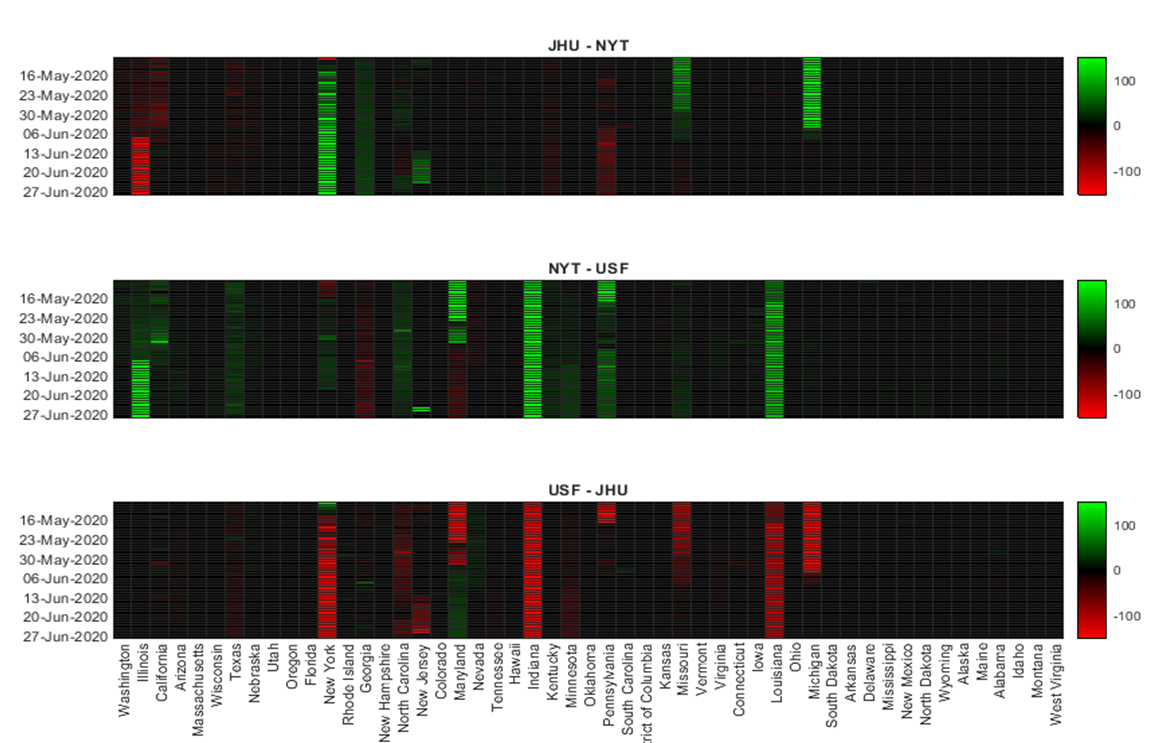} 
\caption{Pairwise differences among the three datasets. Brighter green denotes a higher magnitude positive difference, and brighter red denotes a higher magnitude negative difference. Black represents no difference}
\label{fig:pairwise_diff}
\end{subfigure}
\begin{subfigure}{\textwidth}\centering
\includegraphics[width=0.7\linewidth]{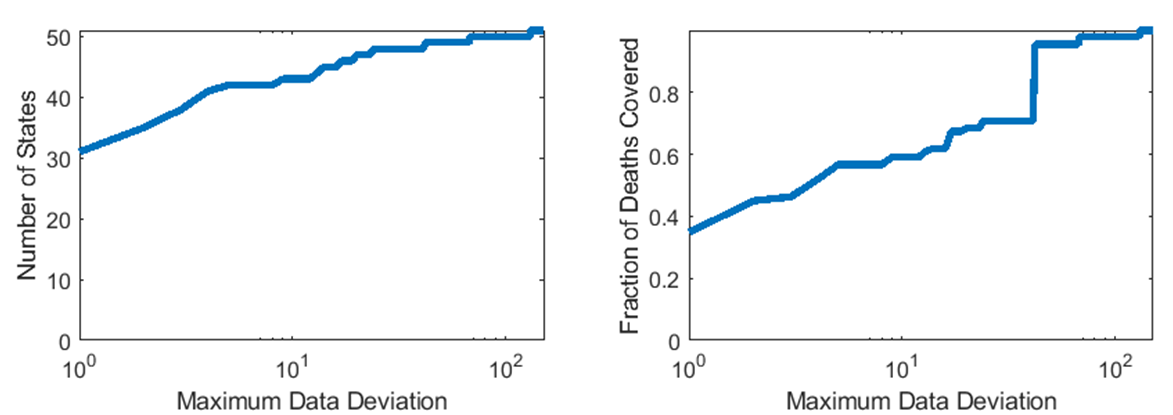}
\caption{Cumulative number of states and fraction of total deaths among states with increasing deviations in the datasets.}
\label{fig:data_range}
\end{subfigure}
\caption{Discrepancies among the three data sources.}
\label{fig:data_disc}
\end{figure*}
We use three sources of data for positive cases and number of deaths in the US states.
\begin{itemize}
    \item JHU: The JHU CSSE COVID19 dataset~\cite{JHUdata}. This dataset is used by YYG\_ParamSearch and Covid19Sim\_Simulator.
    \item NYT: The New York Times dataset~\cite{NYTdata}. This dataset is used by UCLA\_SuEIR, COVIDActNow\_SEIR\_CAN, and IowaStateLW\_STEM.
    \item USF: The US Facts dataset~\cite{USFdata}. This dataset is used by JHU\_IDD\_CovidSP and CU\_select.
\end{itemize}
The three data sources have some inconsistencies as shown in Figure~\ref{fig:pairwise_diff} as three heatmaps obtained by subtracting the data of NYT from JHU (JHU $-$ NYT), USF from NYT (NYT $-$ USF), and JHU from USF (USF$-$JHU).
The y-axis represents the days of the evaluation period and the x-axis represents the states. 
%Brighter green denotes a higher magnitude positive difference, and brighter red denotes a higher magnitude negative difference.
It can be observed that the discrepancy is limited to few states, but it can be significant -- over 150 deaths. Due to such inconsistencies, we present separate evaluation for methods using different data sources.
Figure~\ref{fig:data_range} shows the number of states and deaths covered by various ranges of deviations (absolute value) in data. For a fixed value of the deviation $d$, we pick the states such that the range of the number of deaths for that state at any given point in the evaluation period is less than $d$.
We observe that all three datasets are consistent across 31 states ($d < 1$) that approximately cover approximately 33.7\% of total deaths at the end of our evaluation period (June 28, 2020). For these 31 states, we are able to compare all the methods together.

\begin{table*}[!h]
    \centering
    \caption{Runtimes for various steps in generating learning and forecasting using our approach.}
    \begin{tabular}{|l|c|c|c|}
    \hline
        Step &	US states &	184 countries &	~3000 counties \\
        \hline
Hyper-parameter selection &	2.80 s	& 10.46 s &	2.81 s \\
Reported cases param learning &	0.17 s &	0.55 s &	29.01 s \\
Reported cases forecasts (100 days) &	0.07 s &	0.33 s &	43.46 s \\
Deaths param learning &	0.09 s	& 0.29 s	& 8.37 s \\
Death forecasts (100 days) &	0.08 s &	0.20 s &	17.38 s \\
Total &	3.18 s	& 11.83 s	& 101.03 s \\
\hline
    \end{tabular}
    \label{tab:runtimes}
\end{table*}

\subsection{Runtime}

 We implemented our approach in Matlab R2020a on an Intel(R) Core(TM) i3-3220, 3.30GHz CPU (2 cores) running Windows 10 with 8GB RAM. 
 Since our approach reduces to fitting two linear functions, the parameters can be learned very quickly. Table~\ref{tab:runtimes} shows the runtimes of all the steps involved in generating forecasts -- hyper-parameter tuning, parameter learning for reported cases and deaths, and forecasting reported cases and deaths for 100 days. For all the states in the US, the entire process takes 3.18s. This enables extremely fast scenario analysis. We do not include hyper-parameter tuning for death forecasting, as so far, we have the hyper-parameters fixed, and did not see a significant improvement by identifying new hyper-parameters every time.
 To demonstrate scalability of our approach we have also presented the runtimes for global forecasts for 184 countries, and that for more than 3000 US counties. Our approach takes 11.83s to for the whole process for 184 countries and 101.03s for the US counties. Note that we do not attempt to learn different hyper-parameters for different counties. Instead, we pick the hyper-parameters corresponding to the state that county belongs to. We have observed that picking hyper-parameters independently for regions with small number of infections (which is the case with most counties) can lead to over-fitting.

\begin{table*}[!h]
    \centering
    \caption{Comparison of all models based on average RMSE}
    \begin{tabular}{|c|c|c|c|c|c|c|c|c|}
    \hline
    & \multicolumn{4}{|c|}{Daily RMSE} & \multicolumn{4}{|c|}{Weekly RMSE}\\
    \hline
        Methods & JHU & NYT & USF & No Conflict & JHU & NYT & USF & No Conflict\\
        \hline
        SIkJa & \textbf{23.634} &\textbf{22.01} & \textbf{22.97} &  48.74 & \textbf{75.22} & 74.21& \textbf{78.48} & \textbf{139.23}\\
        YYG\_ParamSearch & 26.67 & &  &  51.54 & 84.42 & & & 156.83\\
        Covid19Sim\_Simulator & 27.82 & & &  65.92 & 105.63 & & & 218.99\\
        UCLA\_SuEIR &  & 22.97 &  &  \textbf{48.13} & & \textbf{73.80} & & 140.84\\
        CovidActNow\_SEIR\_CAN & & 33.08 & &  81.57 & & 130.29& & 256.03\\
        IowaStateLW\_STEM &  & 35.41 &  &  65.31 & & 133.81  & & 209.48\\
        CU\_select &  & & 32.36 & 55.05 & & &120.67 & 175.71\\
        JHU\_IDD\_CovidSP &  & & 48.97 &  67.04 & & & 185.71 & 254.20\\
        \hline
    \end{tabular}
    \label{tab:comparison}
\end{table*}

\subsection{Evaluation Metric}

We evaluate the incident (new) death forecasts using the Root Mean Squared Error (RMSE). We consider two windows of forecasts, daily and weekly. The provided forecasts are daily which can be aggregated over a week to evaluate weekly forecasts. For a given date of forecast release, we compare the true incident deaths to the forecasted deaths over the next 14 days, i.e, over $T = 2$ points for weekly evaluation and over $T=14$ points for daily evaluation. The RMSE is calculated separately for each state and averaged to get the final error. Mathematically, for a given set of states $S$, we evaluate
\begin{equation}
    RMSE = \frac{1}{|S|}\sum_{s\in S} \sqrt{\frac{\sum_\tau (D_\tau - \hat{D}_\tau)^2}{T}}\,.
\end{equation}
$D_\tau$ denotes the forecasted death in the period $\tau$ (day/week) and $\hat{D}_\tau$ denotes the observed incident deaths.

\subsection{Results}
Table~\ref{tab:comparison} provides a comparison of all the methods based on RMSE measured daily and weekly, averaged ovber the evaluation period. Our approach (SIkJ$\alpha$) produces the best forecasts in all the datasets based on daily RMSE. For RMSE calculated weekly, UCLA\_SuEIR performs slightly better in case  of NYT dataset. For other datasets, our approach tops the list. When considering the 31 states which do not show any conflict among the three datasets, our approach performs the best based on weekly RMSE, while UCLA\_SuEIR outperforms our approach by a close margin. Overall, in most cases, our approach produces the best RMSE, and in other cases it is close to the performance of best performing baseline.

\begin{figure*}[!htbp]
\centering
\includegraphics[width=\textwidth]{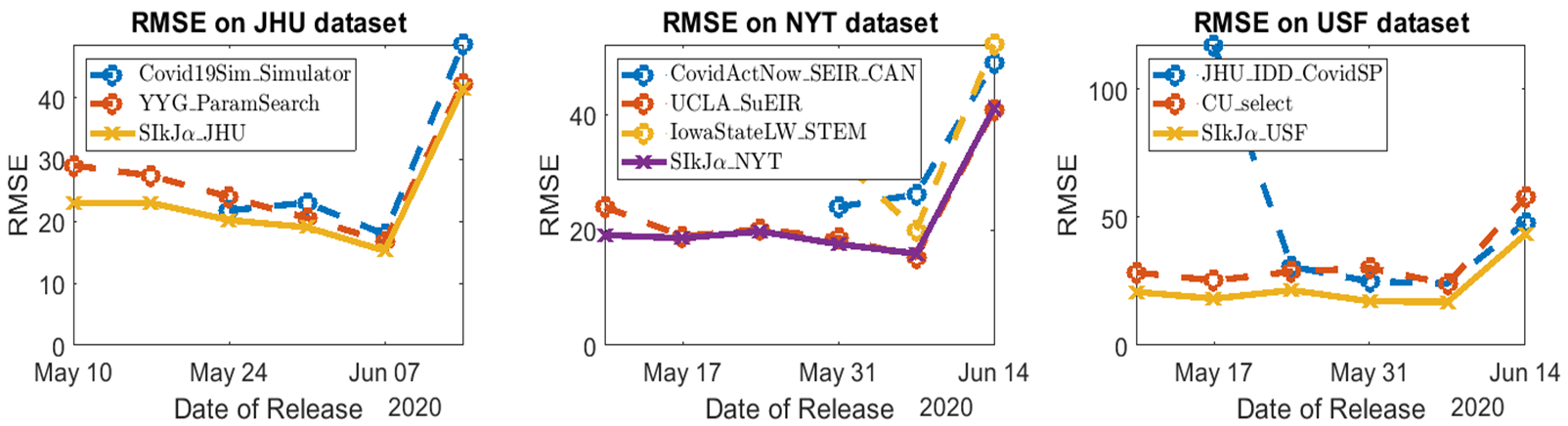} 
\caption{Comparison of all the methods separated by their respective datasets, based on RMSE measured daily.}
\label{fig:RMSE_all}
\end{figure*}

Figures~\ref{fig:RMSE_all} shows the performance of all the forecasts released weekly based on daily RMSE aggregated over two weeks. Note that a higher RMSE on one week compared to another does not imply worse performance as higher ground truth may lead to higher RMSE. These plots should be assessed based on relative performance on every day of forecast release rather than absolute numbers across the weeks. We observe that our approach SIkJ$\alpha$consistently performs well over the evaluation period on all three datasets.

\begin{figure*}[!htbp]
\centering
\includegraphics[width=\textwidth]{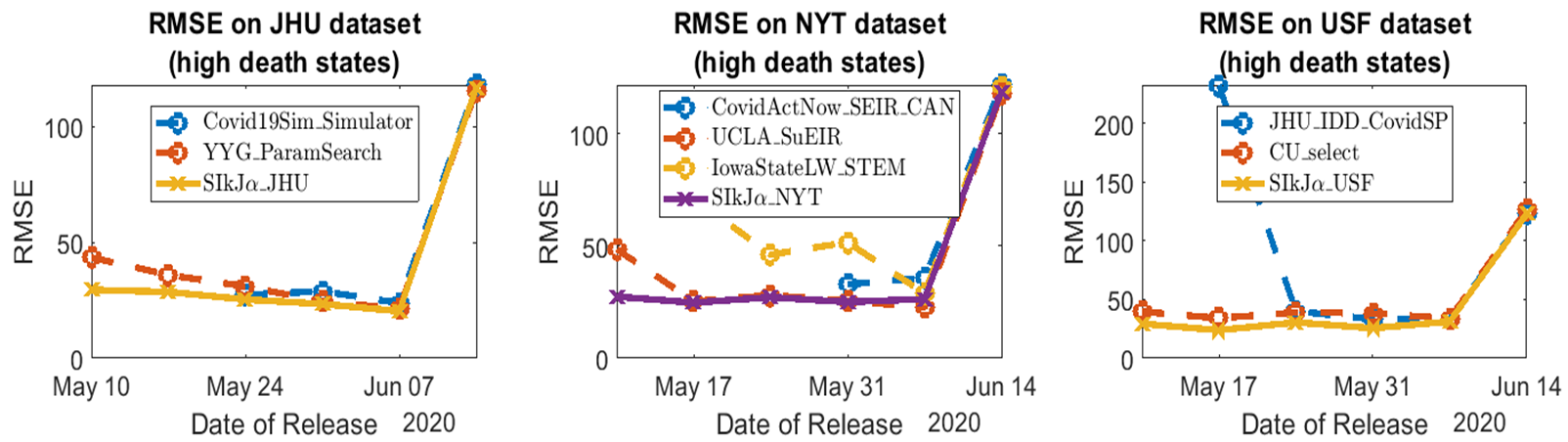} 
\caption{Comparison of all the methods on high-death states separated by their respective datasets, based on RMSE measured daily.}
\label{fig:RMSE_high}
\end{figure*}

\begin{figure*}[!htbp]
\centering
\includegraphics[width=\textwidth]{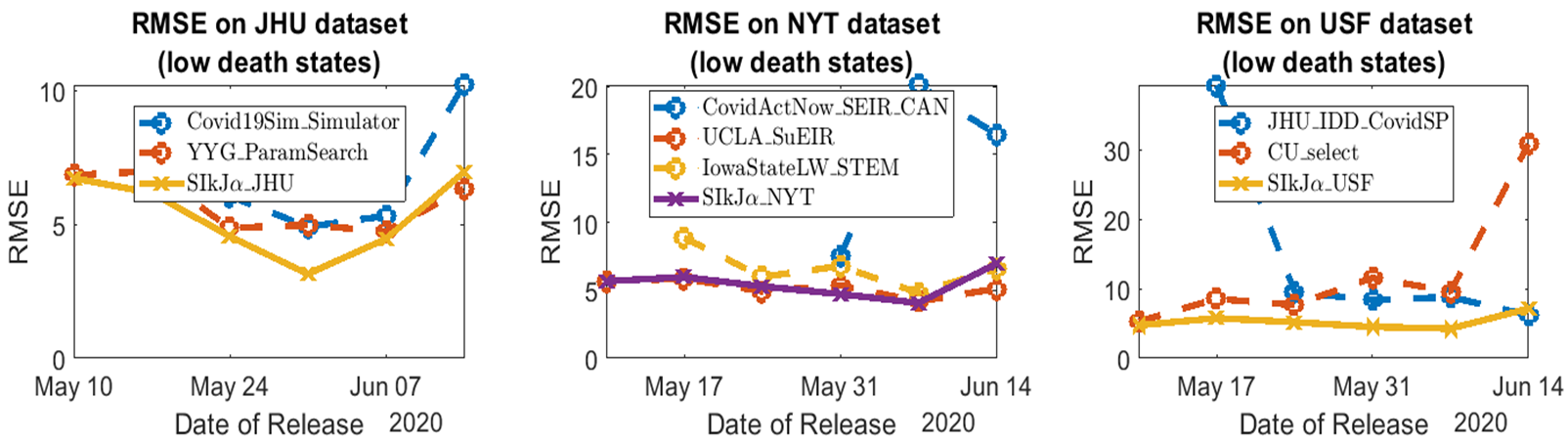} 
\caption{Comparison of all the methods on low-death states separated by their respective datasets, based on RMSE measured daily.}
\label{fig:RMSE_low}
\end{figure*}

To observe the effect of data on the performance, we split the dataset into two halves - high death states consisting of the top 25 states with most deaths at the start of the evaluation, and low death states constituting the rest of the states. The results are shown in Figures~\ref{fig:RMSE_high} and~\ref{fig:RMSE_low} . We observe that the relative performance all the methods are closer to each other on high death states compared to low death states. This is expected as states with higher deaths are likely to average out noise/anomalies, making it easier to train models and forecast.

\begin{figure*}[!htbp]
\centering
\begin{subfigure}{0.49\textwidth}
\includegraphics[width=\linewidth]{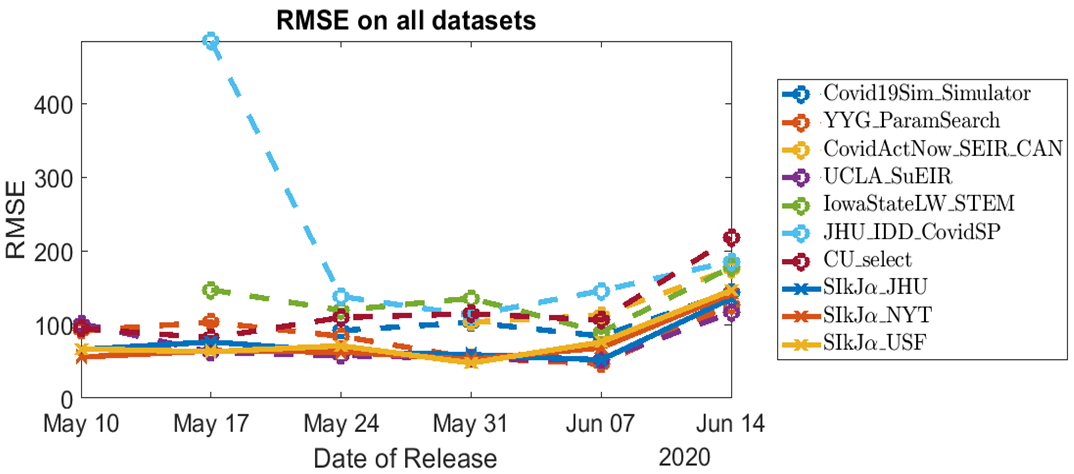} 
\label{fig:all_graph_weekly}
\end{subfigure}
\begin{subfigure}{0.49\textwidth}
\includegraphics[width=\linewidth]{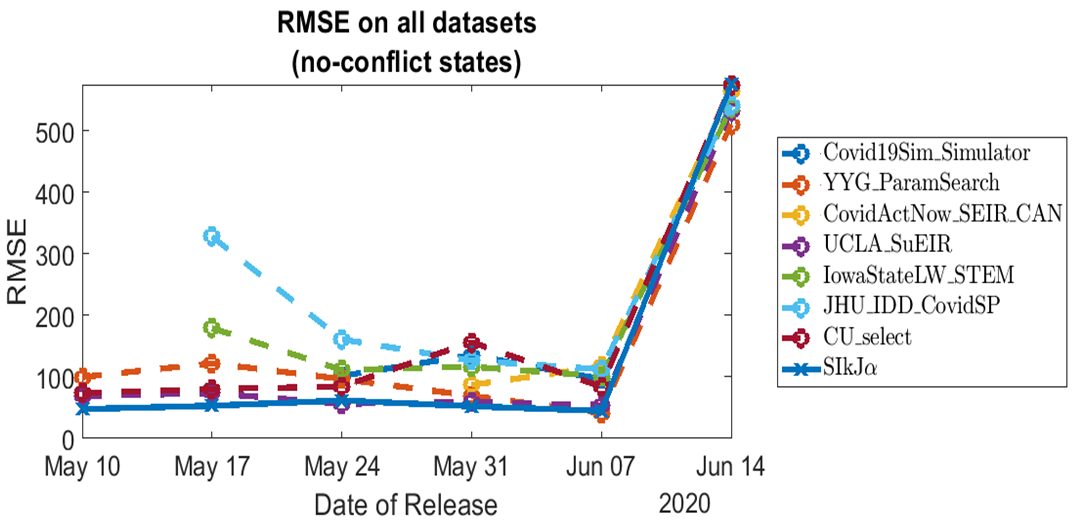}
\label{fig:all_graph_no_conflict_weekly}
\end{subfigure}
\caption{Comparison of the methods on all datasets on all states and the states with no-conflict, based on RMSE measured weekly.}
\label{fig:all_graph_no_conflict}
\end{figure*}

We also evaluated all the methods on the 31 states for which there is no conflict among the three datasets (see Figure~\ref{fig:all_graph_no_conflict}). We compare this against the errors obtained by all the methods on the three datasets calculated on each forecast day. We observe that the relative errors of the methods are closer to each other when considering the states with no conflict. This suggests that the inconsistency in reporting may have introduced additional noise making it difficult to train the models. 

We repeated the above experiments for the forecasts released weekly based on weekly RMSE aggregated over two weeks. We observe similar trends in all the figures as noted above, with the exception of YYG\_ParamSearch and UCLA\_SuEIR performing slightly better than our approach on some forecasts. Our online interactive website has a dedicated page to update and show the comparison our approach with the baselines\footnote{\url{https://scc-usc.github.io/ReCOVER-COVID-19/\#/leaderboard}}.

\begin{figure*}[!htbp]
\centering
\begin{subfigure}{0.49\textwidth}
\includegraphics[width=\linewidth]{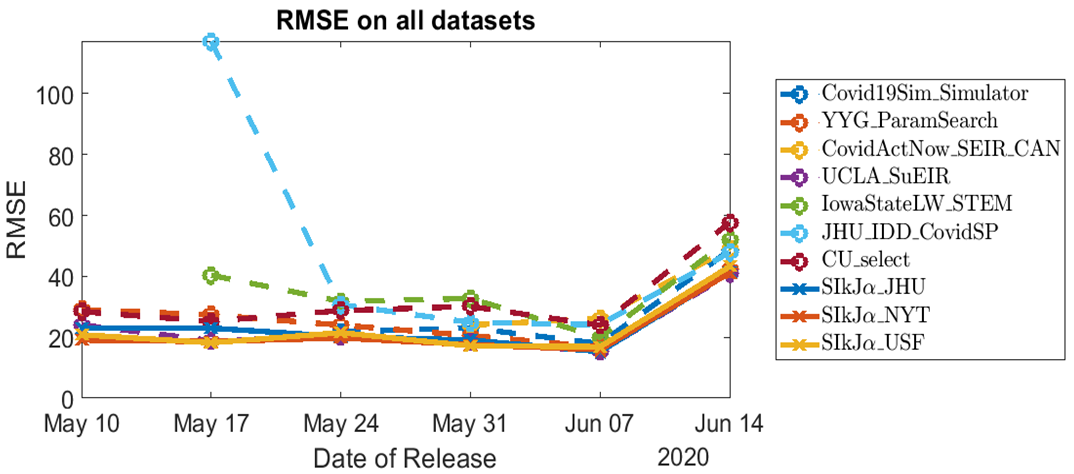} 
\label{fig:all_graphs}
\end{subfigure}
\begin{subfigure}{0.49\textwidth}
\includegraphics[width=\linewidth]{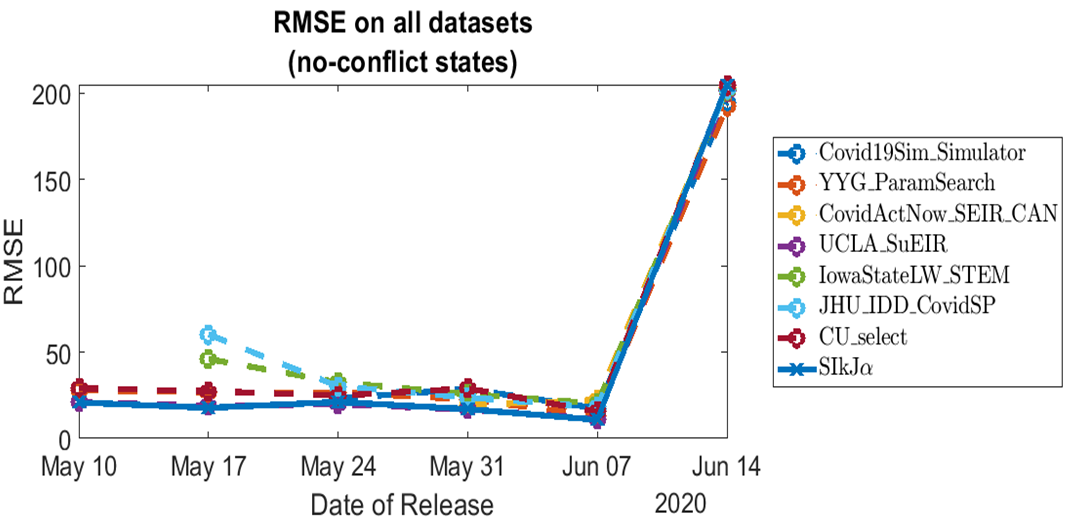}
\label{fig:all_graphs_no_conflict_daily}
\end{subfigure}
\caption{Comparison of the methods on all datasets on all states and the states with no-conflict, based on RMSE measured daily.}
\label{fig:all_graphs_no_conflict}
\end{figure*}

\begin{figure*}[!htbp]
\centering
\includegraphics[width=\textwidth]{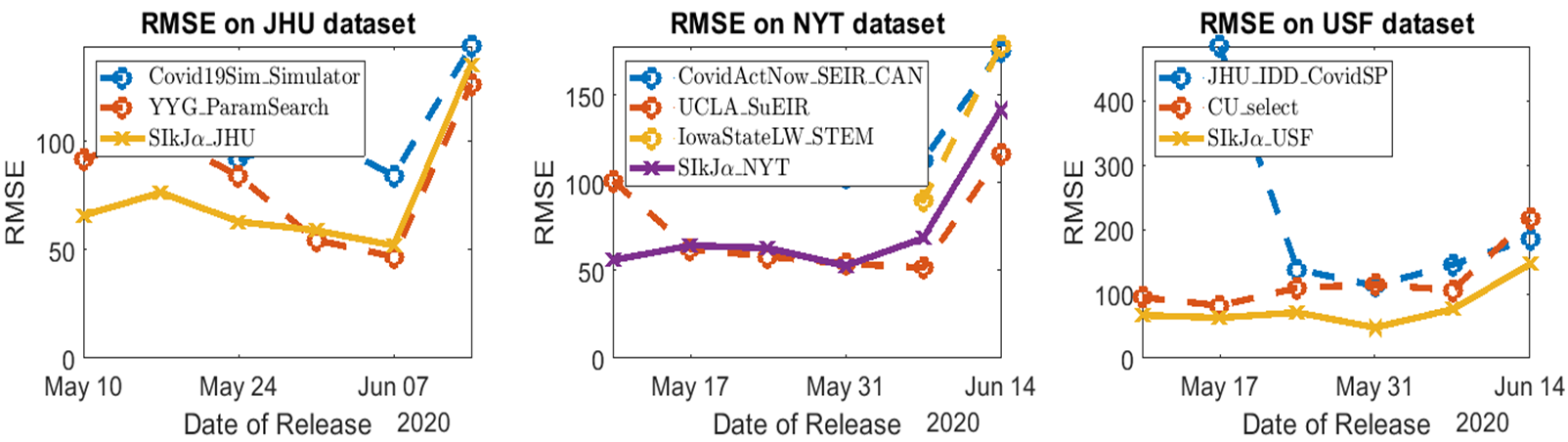} 
\caption{Comparison of all the methods on high-death states separated by their respective datasets, based on RMSE measured weekly.}
\label{fig:RMSE_all_weekly}
\end{figure*}

\begin{figure*}[!htbp]
\centering
\includegraphics[width=\textwidth]{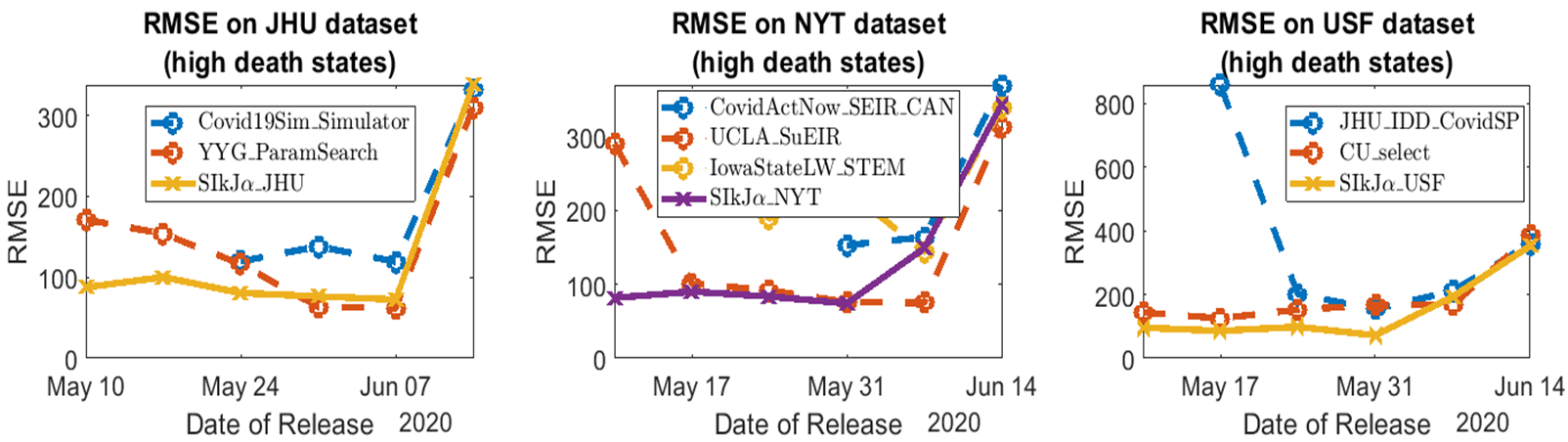} 
\caption{Comparison of all the methods on high-death states separated by their respective datasets, based on RMSE measured weekly.}
\label{RMSE_high_weekly}
\end{figure*}

\begin{figure*}[!htbp]
\centering
\includegraphics[width=\textwidth]{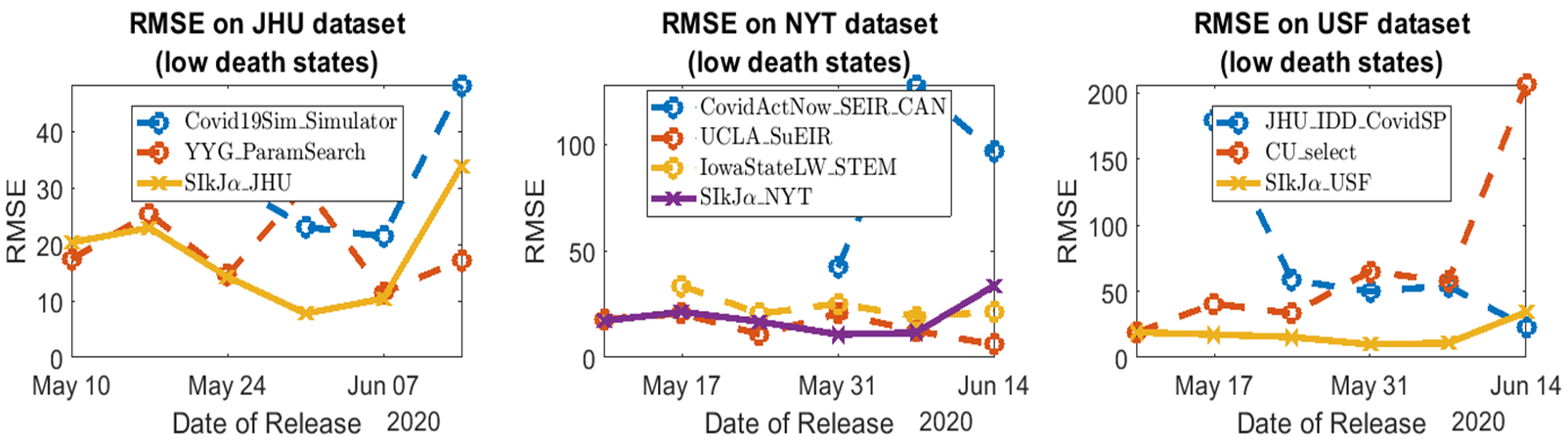} 
\caption{Comparison of all the methods on low-death states separated by their respective datasets, based on RMSE measured weekly.}
\label{RMSE_low_weekly}
\end{figure*}

\section{Discussions}
\paragraph{Reproduction Number.} Based on the learned values of $\beta_i^{(p)}$ at a time $t$ for a region $p$, we can identify the dynamic reproduction number~\cite{huang2020dynamic} of the epidemic in that region, defined as the number of new infections created by one infected individual in the remaining susceptible population. Suppose, the individual is infected at time $t$, then according to Equations~\ref{eqn:new_infections}, they remain active for $kJ$ time steps. From $t+1$ to $t+J$ they will contribute to new infections at the rate of $\beta_1^{(p)}$, from $t+J+1$ to $t+2J$ at the rate of $\beta_2^{(p)}$, and so on. Therefore, the total infections created by one individual is:
\begin{align}
    \mathcal{R}_t^{(p)} &= J\sum_i \beta_i^{(p)} \sum_{j=1}^{J} \left( 1 - \frac{I_{\tau + (i-1)J+j}}{(1-\rho^{(p)}) N^{(p)}}
    \right)\,\nonumber\\
    &\approx \left(1 - \frac{R_t^{(p)}}{\gbar N} \right)J\sum_i \beta_i^{(p)}
\end{align}
Note that if we assume that the new reported cases in Equation~\ref{eqn:report_final} in the last $kJ$ time steps are roughly equal, then
\begin{equation}
    \Delta R_t^{(p)} = \left(1 - \frac{R_t^{(p)}}{\gbar N} \right)J\sum_i \beta_i^{(p)} \Delta R_{t-1}^{(p)} =  (\mathcal{R}_t^{(p)}) \Delta R_{t-1}^{(p)}\,,
\end{equation}
which suggests that the traditional interpretation of reproduction number applies to our model as well -- if $\mathcal{R}_t^{(p)} < 1$ the disease will die out, as it will create fewer new infections. The basic reproduction number, that assumes the entire population to be susceptible, is $\mathcal{R}_0^{(p)} = J\sum_i \beta_i$.

\paragraph{Model-based Case Fatality Rate.} Case fatality rate is given by the fraction of reported positive cases that end up in death. A simple way of calculating this is by taking the ratio of total deaths and total reported cases. However, this does not consider the lag between reporting positive and then dying, and also does not account for recent changes in the dynamics. The learned values of $\thebar_i^{(p)}$ can be used to compute model-based case fatality rate (MCFR). According to Equation~\ref{eqn:death_final}, an individual reported positive at time $t$ dies on each of the time steps among $t+1$ to $t+J$ with probability $\thebar_i^{(p)}$, $t+J+1$ to $t+2J$, and so on. Therefore, MCFR can be calculated as expected number of new deaths created by one reported case as
\begin{equation}
    MCFR_t = J\sum_i \thebar_i^{(p)}\,.
\end{equation}

The weekly updated reproduction numbers and model-based case fatality rates for all US states and all the countries can be found on our webpage\footnote{\url{https://scc-usc.github.io/ReCOVER-COVID-19/\#/score}}.

\section*{Acknowledgments}
This work was supported by National Science Foundation Award No. 2027007.

\bibliographystyle{plain}
\bibliography{refs}

\end{document}